\documentclass[aps, prd, twocolumn, lengthcheck, superscriptaddress, showpacs,
nofootinbib]{revtex4-1}

\usepackage{epsfig}
\usepackage[usenames]{color}
\usepackage{graphicx}
\usepackage{amsmath}
\usepackage{epstopdf}
\usepackage{sidecap}
\usepackage{rotating} 
\usepackage{subfigure}
\usepackage{multirow}

\def\bi{\bibitem}

\def\be{\begin{eqnarray}}\def\ee{\end{eqnarray}}
\def\lsim{\mathrel{\rlap{\lower3pt\hbox{\hskip1pt$\sim$}}
     \raise1pt\hbox{$<$}}} 
\def\gsim{\mathrel{\rlap{\lower3pt\hbox{\hskip1pt$\sim$}}
     \raise1pt\hbox{$>$}}} 

\allowdisplaybreaks

\begin{document}

\title{Topology change and emergent scale symmetry in compact star matter via gravitational wave detection}

\author{Wen-Cong Yang}
\affiliation{Center for Theoretical Physics and College of Physics, Jilin University, Changchun, 130012, China}

\author{Yong-Liang Ma}
\email{ylma@ucas.ac.cn}
\affiliation{School of Fundamental Physics and Mathematical Sciences,
Hangzhou Institute for Advanced Study, UCAS, Hangzhou, 310024, China}
\affiliation{International Center for Theoretical Physics Asia-Pacific (ICTP-AP) (Beijing/Hangzhou), UCAS, Beijing 100190, China}

\author{Yue-Liang Wu}
\email{ylwu@ucas.ac.cn}
\affiliation{School of Fundamental Physics and Mathematical Sciences,
Hangzhou Institute for Advanced Study, UCAS, Hangzhou, 310024, China}
\affiliation{International Center for Theoretical Physics Asia-Pacific (ICTP-AP) (Beijing/Hangzhou), UCAS, Beijing 100190, China}
\affiliation{Institute of Theoretical Physics, Chinese Academy of Sciences, Beijing, 100190, China}
\affiliation{School of Physical Sciences, University of Chinese Academy of Sciences, Beijing, 100049, China}

\date{\today}

\begin{abstract}
Topological structure has been extensively studied and confirmed in highly correlated condensed matter physics. We explore the gravitational waves emitted from binary neutron star mergers using the pseudoconformal model for dense nuclear matter for compact stars. This model considers the topology change and the possible emergent scale symmetry and satisfies all the constraints from astrophysics. We find that the location of the topology change affects gravitational waves dramatically owing to its effect on the equation of state. In addition, the effect of this location on the waveforms of the gravitational waves is within the ability of the on-going and up-coming facilities for detecting gravitational waves, thus suggesting a possible way to measure the topology structure in nuclear physics.
\\
\textbf{Keywords: } Topology change; emergent symmetry; compact star matter; gravitational wave.
\end{abstract}

\pacs{04.30.Db, 12.39.Fe, 26.60.+c, 97.60.Jd}

\maketitle

\section{ Introduction}

The nature of strongly interacting matter at a high baryon number density is one of the outstanding open problems in both nuclear physics and astrophysics. What are the symmetry patterns involved in this region? What are the constituents at high density relevant to the cores of the compact stars? Do any novel phenomena occur inside massive compact stars? For some discussions on these aspects, we suggest, e.g., Refs.~\cite{Holt:2014hma,Baym:2017whm,McLerran:2018hbz,Ma:2019ery,Strangness} and some relevant references therein.  At this moment, these puzzles can neither be clarified by fundamental quantum chromodynamics (QCD)---even using lattice simulations---nor be judged by the terrestrial experiments.

The detection of gravitational waves (GWs) from neutron star mergers has ushered in a new era of nuclear physics~\cite{TheLIGOScientific:2017qsa}. The properties of the GWs emitted during the merger, including the amplitude and frequency, are closely related to the star masses and star structures. In general, the GWs carry much information of the inner structure of the stars, including the equation of state (EoS) of the nuclear matter and the nature of the strongly interacting baryonic matter. In addition, the GWs emitted in the postmerger stage decay rapidly and provide the information about the baby star mass and spin that also depend on the EoS of the nuclear matter.

In this study, we examine the GWs emitted from neutron star mergers using a conceptually novel approach to dense nuclear matter anchored on some symmetries emerging in the high-density region as well as a particular topological structure of baryonic matter embodying both nucleonic and quarkonic properties~\cite{MR-SV,PCM} (for a systematical review, see \cite{Ma:2019ery}). This model is found to satisfy all the constraints determined from the terrestrial experiments as well as from astrophysical predictions that the matter in the core of massive compact stars satisfies the pseudoconformal (PC) velocity $v_s^2/c^2 \simeq 1/3$ but is still composed of confined quasifermions with fractional baryon charge~\cite{Ma:2020hno}. We call this model the pseudoconformal model (PCM). This is in stark contrast to the widely accepted belief that the conformal sound velocity appears at $ \sim 100n_0$ where perturbative QCD is applicable and deconfined quarks figure in ~\cite{Tews:2018kmu}.

\section{Topology change and emergent symmetry}

In this section, we first summarize the key points of the model used in this work. For the details of the model, see Refs.\cite{Ma:2019ery,MR-book}. One way to approach the nuclear many-body problem is to use the effective theory incorporating mesons only and to regard baryons as topology objects carrying winding number one---skyrmions---and to put skyrmions onto a certain crystal lattice. A robust conclusion found in this approach is that there is a topology change corresponding to the skyrmion--half-skyrmion transition with the half-skyrmion carrying a winding number-1/2~\cite{MR_SkyrRev}. By the Cheshire Cat Principle expounded in \cite{Ma:2019ery},  the location of the topology change $n_{1/2}$ is found to be confined in the range $(2 \lsim n_{1/2}/n_0 < 4)$  to capture the putative hadron--quark continuity.

This topology change has several important impacts on the EoS for densities $n > n_{1/2}$ \cite{Ma:2019ery}. Some of them relevant to the present work are as follows: (i) The quark condensate vanishes globally but not locally with nonvanishing and nearly density-independent pion and dilaton---will be introduced later---decay constants $f_\pi \sim f_\chi \neq 0$. (ii) The baryon mass becomes a density-independent constant with a magnitude $m_0\simeq (0.6 - 0.9)m_N$, indicating the emergence of the parity-doubling structure of nucleons. (iii) The hidden gauge coupling associated with the $\rho$ meson mass, to be introduced later, starts to drop and flows to zero at the vector manifestation fixed point~\cite{Harada:2003jx}, and therefore, the vector meson becomes massless, and the hidden gauge symmetry emerges.

Although the skyrmion crystal approach can provide some robust and valuable information on the possible topology structure of nuclear matter beyond saturation density $n_0$, so far it has not been put into practice owing to the lack of systematic and reliable formulation of the many-body dynamics. In practice, we resort to the effective theory in which baryons are included as explicit degrees of freedom and incorporate the robust conclusion from the skyrmion crystal---the density dependence of physical quantities and the topology change---to the parameters in the baryonic effective theory. For a systematic review, please refer to \cite{Ma:2019ery}.

To access the density region relevant to compact stars, i.e., $\lesssim 10 n_0$, the lowest vector mesons $V = (\rho,\omega)$ and the scalar meson $f_0(500)$ should be included in addition to pions and nucleons (denoted as G$n$EFT) because the power counting in the Fermi momentum using in the effective theory breaks in such a high-density region. The vector mesons $V = (\rho,\omega)$ are regarded as the dynamical fields of hidden local symmetry (HLS) ~\cite{Harada:2003jx}, and the scalar $\chi$---regarded as $f_0(500)$ in the particle data booklet---as the Nambu--Goldstone boson of hidden scale symmetry\cite{CT}. Both symmetries are hidden in the matter-free space as well as the nuclear matter around saturation density and are expected to emerge in a superdense region relevant to compact stars. The appearance of these symmetries in dense medium has been discussed in~\cite{PCM,Ma:2019ery,Ma:2020nih}.

Now, the key strategy in our approach is to map the robust conclusion obtained from the skyrmion lattice approach to the parameters of the G$n$EFT. The density effect enters the system from both the intrinsic density dependence (IDD) inherited from fundamental QCD~\cite{br91} and the nuclear correlations in nuclear matter (denoted as DD$_{\rm induced}$). Among a variety of outcomes, IDD gives a reasonable explanation of the $g_A$ quenching problem in the superallowed Gamow--Teller transitions~\cite{LMRgA,MaRhogAPRL}. Owing to the topology change, the IDD is divided into two density regimes R-I and R-II that are delineated by the topology change density $n_{1/2}$. In R-I, the IDDs are fixed by only one parameter that gives the density dependence of the in-medium pion decay constant as determined from deeply bound pionic atoms. While in R-II, the topology together with the assumed high-density properties of hidden local symmetry and (hidden) scale symmetry determine the variation of the EoS should with density. The pseudoconformal structure results from this property in R-II.

The most striking prediction of the model is the precocious onset of the conformal sound speed $v_s^2/c^2=1/3$ in massive stars with matter density $n \geq n_{1/2}$, though the trace of the energy--momentum tensor is nonzero. In the cores of massive neutron stars, the sound speed satisfies $v_s^2/c^2=1/3$, and the polytropic index $\gamma< 1.75$, which is smaller than the minimum value found in the hadronic models. This is in contrast to the standard scenario~\cite{Annala_NP}, while the deconfined quarks do not figure in~\cite{Ma:2020hno}.

In the present work, we study the effect of the topology change on the waveform of the GWs emitted from the merger of the BNSs by taking the typical values $n_{1/2}/n_0 = 2.0$ and $3.0$, which are within the constraints from astrophysics. In the simulation, we consider the equal mass BNS mergers with star masses $1.5M_\odot$ and $1.7M_\odot$. We find that a detector with a resolution $O(1)$ms can distinguish the location of $n_{1/2}$. This is within the ability of the on-going and up-coming GW detection facilities; see~\cite{GWDetection} and the references therein. {\it To the best of our knowledge, this is the first analysis of the topology effect on the direct measurable in the strongly interacting baryonic matter, though it has been widely studied and confirmed in condensed matter physics, both theoretically and experimentally~\cite{Tong:2016kpv}.} Therefore, this work is expected to usher in a possible paradigm shift in both nuclear physics and astrophysics.

\section{Pseudoconformal model and star properties}

The EoS of the PCM, which is used in the BNS merger, is calculated by the $V_{lowk}$ renormalization-group ($V_{lowk}$-RG) approach~\cite{Vlowk}. At low density up to $\lesssim n_{1/2}$, the nuclear matter properties can be reproduced well~\cite{Ma:2020hno} with a suitable choice of the scaling parameters in IDD. Above $n_{1/2}$, the topology change affects the EoS drastically and consequently the star properties.

Based on these considerations, the only parameter that cannot be constrained by the theory is the location of the topology change $n_{1/2}$. Previously, using some indirect information from astrophysics, we constrained $2.0 \lesssim n_{1/2}/n_0 < 4.0 $~\cite{Ma:2019ery}. In this work, we devote ourselves to the effect of $n_{1/2}$ on the direct observables in astrophysics---the GWs emitted from a BNS merger.

\begin{figure}[tbp]
	\centering
	\subfigure[]{
		\includegraphics[width=1.55in]{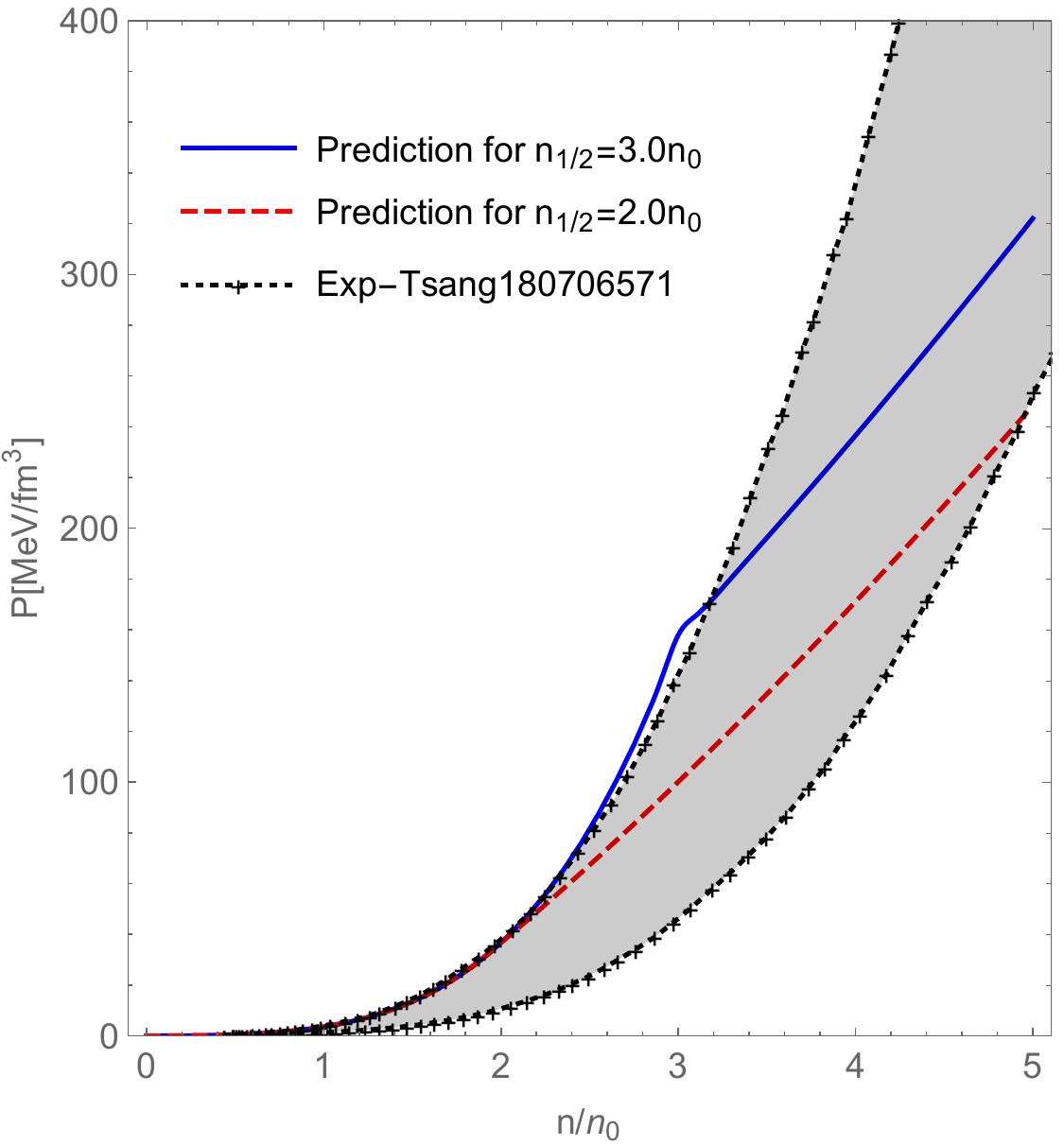}
		\label{fig:EoS}
	}
	\subfigure[]{
		\includegraphics[width=1.55in]{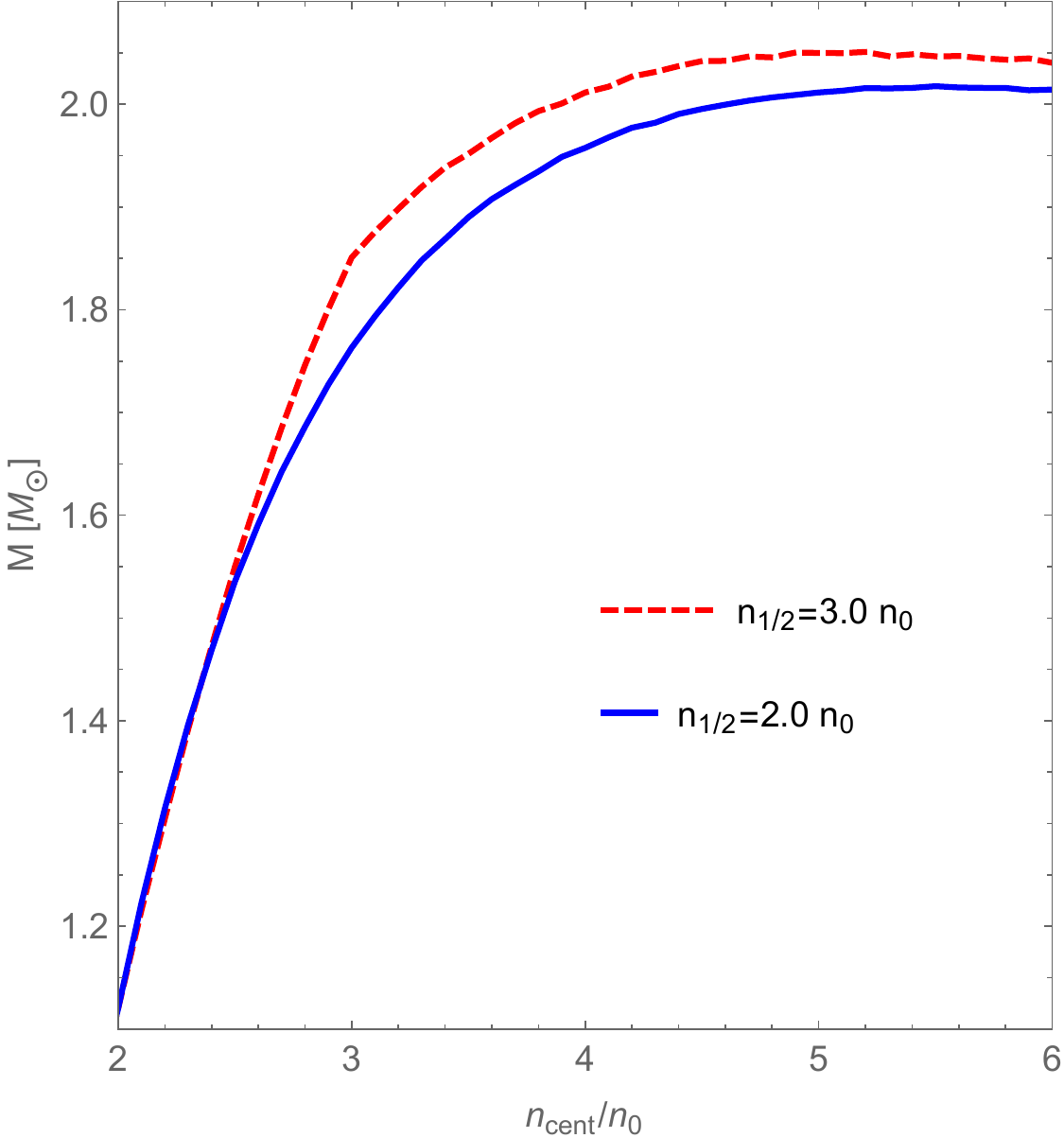}
		\label{fig:Star}
	}
	\caption{Equation of state (a) and star properties (b) from the pseudoconformal model (PCM). The available experimental bound (shaded) is given in~\cite{Tsang:2018kqj}.}
	\label{fig:EoSStar}
\end{figure}

In Fig.~\ref{fig:EoSStar}, the EoS and star mass are plotted as functions of central density. From Fig.~\ref{fig:EoS}, one can see that the predicted EoSs satisfy the constraint from astrophysics and the bigger $n_{1/2}$, the stiffer is the EoS. Further, Fig.~\ref{fig:Star} tells us that the topology change affects the central density of a heavier star more significantly.

The star properties relevant to the present work are listed in Table~\ref{tab:StarProperty}. As pointed above, for a heavier star with a mass $1.7M_\odot$, the star properties are more sensitive to the location of $n_{1/2}$. The sensitivity of the central density indicates that the core of a massive star made of pseudoconformal matter---not the deconfined quark---is bigger size~\cite{Ma:2020hno}. However, the global properties of the lighter star with mass $1.5 M_\odot$, for example, the tidal deformability $\Lambda$ and radius $R$, are not as sensitive to the value of $n_{1/2}$. Hence, it is not easy to pin down the value of $n_{1/2}$ using these values. Unlike the global properties, the central pressure $P_{\rm cent}$ of the light star $1.5M_\odot$ is changed about $O(1\%)$ when $n_{1/2}$ is changed from $2.0n_0$ to $3.0n_0$. Later, we will see that the waveforms of GWs are more sensitive to this difference.

\begin{widetext}
\begin{table*}[!t]
\centering
\caption{Properties of compact stars with different masses and $n_{1/2}/n_0$.}
\label{tab:StarProperty}
\begin{tabular}{c|cc|cc|cc|cc}
\hline
\hline
\multirow{2}{*}{$M/M_\odot$}& \multicolumn{2}{c|}{$n_{cent}/n_0$}&\multicolumn{2}{|c}{ $\Lambda/100$}&\multicolumn{2}{|c}{ $R$/km}&\multicolumn{2}{|c}{ $P_{\rm cent}$/(MeV/fm$^3$)}\cr\cline{2-9}
&$n_{1/2}=2.0$&$n_{1/2}=3.0$&$n_{1/2}=2.0$&$n_{1/2}=3.0$&$n_{1/2}=2.0$&$n_{1/2}=3.0$&$n_{1/2}=2.0$&$n_{1/2}=3.0$
\cr
\hline
1.50 &2.44&2.44&4.35&4.35&12.85&12.85&74.92&74.56\cr\hline
1.70 &2.83&2.72&1.87&1.90&12.83&12.84&103.40&127.23\cr\hline
\hline
\end{tabular}
\end{table*}
\end{widetext}

\section{Binary neutron star merger}

In order to investigate the effect of topology change on the waveform of the GWs emitted from BNS mergers, we perform the numerical relativity simulations by using the Einstein Toolkit~\cite{ETK,ETK2}, which is a community-driven software platform of core computational tools~\cite{Schnetter:2003rb,Baiotti:2004wn,Hawke:2005zw,GRHydro,McLachlan:web,Cactuscode:web} to advance and support research in relativistic astrophysics and gravitational physics. In addition, to generate the initial conditions~\cite{initialData} of irrotational binary systems  (e.g., the initial frequency and angular momentum), the LORENE library~\cite{LORENE} is also needed.

To simulate the processes of BNS mergers using the aforementioned tools, we further parameterize the EoS plotted in Fig.~\ref{fig:EoS} --- with the beta equilibrium correction --- to the piecewise polytrope form~\cite{piecewise poly}. For each piece of matter density $\rho_{i-1}\leq\rho\leq\rho_i$, the pressure and density satisfy the polytropic form
\be
p(\rho) = K_i\rho^{\Gamma_i}.
\ee
As a consequence of continuity, the adiabatic index $\Gamma_i$ and coefficient $K_i$ of the two adiabatic immediate density regions satisfy
\begin{equation}
	K_{i+1} = \frac{p(\rho_i)}{\rho_i^{\Gamma_i +1}}.
\end{equation}
In the present work, we fit the EoS using five polytropic pieces. All the fitted parameters are listed in Table~\ref{table:piecewise}.
\begin{table*}[htbp]
\caption{Parameters in a piecewise polytropic form of the EoS. All variables are expressed in CU.}
\label{table:piecewise}	
\begin{tabular}{ccccccccccc}
\hline
\hline
		$n_{1/2}$ & $K_0$ & $\Gamma_0$ & $\Gamma_1$  & $\Gamma_2$ & $\Gamma_3$ & $\Gamma_4$ & $\rho_1$& $\rho_2$& $\rho_3$& $\rho_4$ \\
		\hline
		$2n_0$& 0.0507 & 1.3138 & 1.6315 & 3.1544 & 2.0611 & 1.6080 & $5.5781\times10^{-5}$ & $2.2637\times10^{-4}$ & $1.0954\times10^{-4}$ & $1.7504\times10^{-3}$\\
		$3n_0$& 0.0406 & 1.2948 & 1.5651 & 2.8001 & 3.1965 & 1.6316 & $4.1820\times10^{-5}$ &
		$1.9772\times10^{-4}$ & $3.4953\times10^{-4}$ & $1.2463\times10^{-3}$\\
\hline
\hline
\end{tabular}

\end{table*}
In the simulation of the BNS merger, we include the thermal effects by supplementing a thermal component on EoS
\be
P_{th} = (\Gamma_{th} - 1)\rho\varepsilon_{th},
\ee
where $\varepsilon_{th}$ is the thermal part of specific internal energy, and thermal adiabatic index $\Gamma_{th}$ can be taken as $\Gamma_{th}=1.8$~\cite{thermal component}.

As stated above, to simulate how the topology change affects BNS mergers, we consider the typical values $n_{1/2} = 2n_0$ and $3n_0$ and neutron star masses $1.5M_\odot$ and $1.7 M_\odot$. In each simulation, we set the initial separation between two star centers as $40$~km and perform the simulation with BSSN-NOK formulation~\cite{BSSN1,BSSN2} of the Einstein equation and PPM reconstruction method~\cite{PPM} of hydrodynamical variables. The resolution of the simulation grid is set as $dx=0.75$ CU (computation unit (CU) takes $c=G = M_\odot =1$) as suggested in Ref.~\cite{resolution}.

\begin{figure}[tbp]
	\centering
	\subfigure[]{
		\includegraphics[width=3.2in]{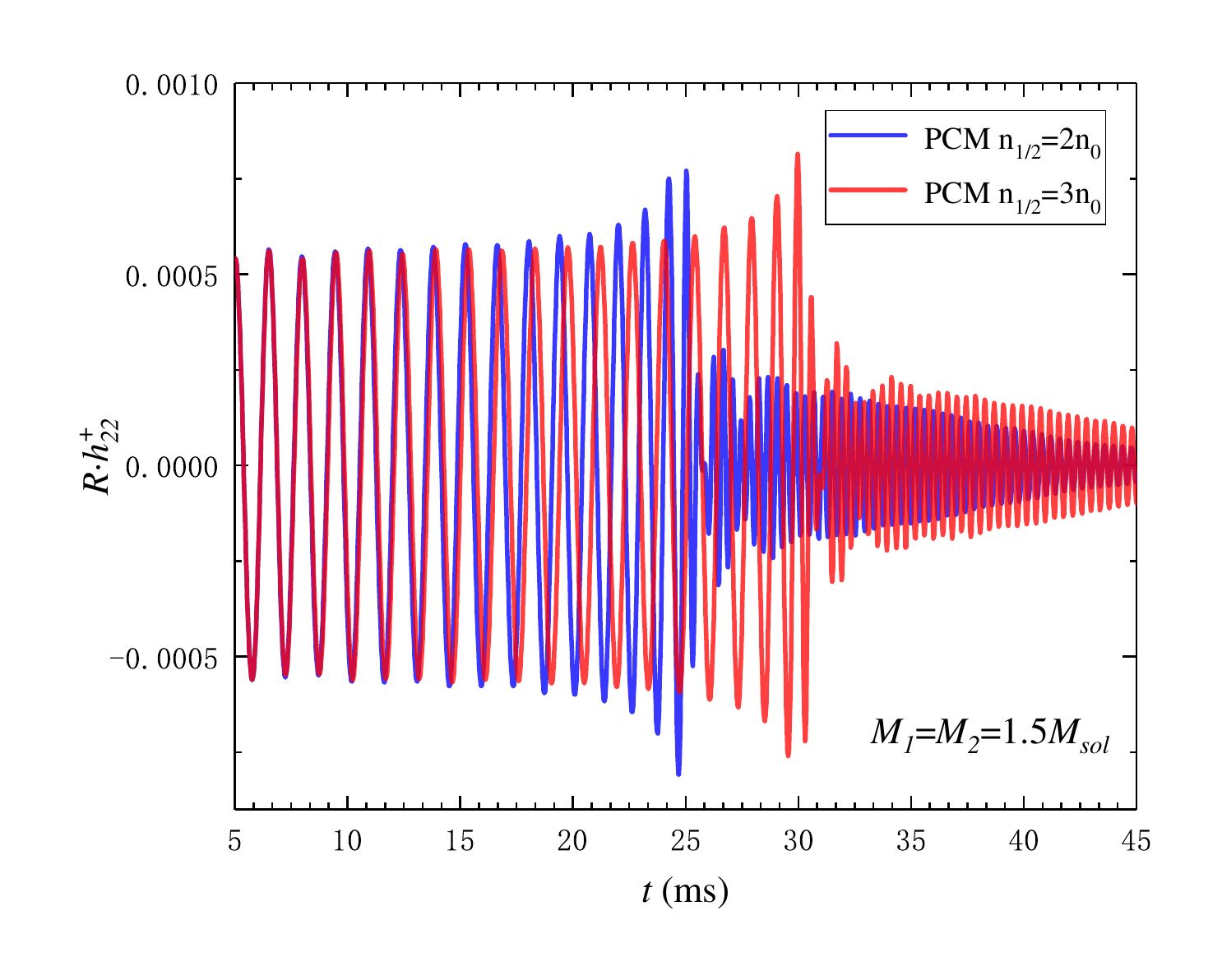}
		\label{fig:GWa}
	}
\\
	\subfigure[]{
		\includegraphics[width=3.2in]{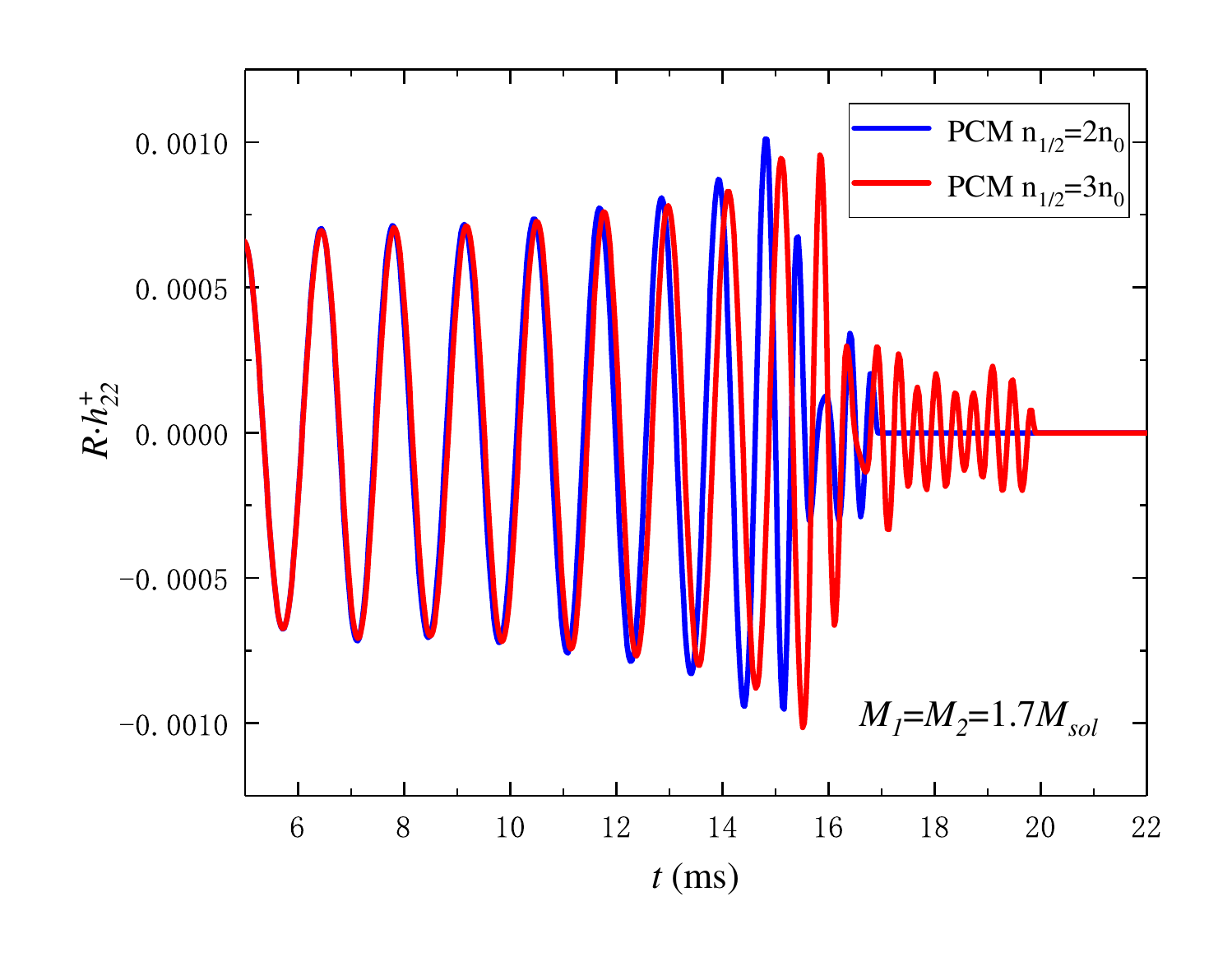}
		\label{fig:GWb}
	}
	\caption{Dominant mode of the GW strain from the BNS star system with equal masses (a) $1.5 M_{\odot}$ stars and (b) $1.7 M_{\odot}$ stars.
All signals are extracted at $R=700$~CU from the center of the binary and are extrapolated to spatial infinity and integrated with an IIR filter provided in Ref.~\cite{IIR}.}
	\label{fig:GW}
\end{figure}

In Fig.~\ref{fig:GW}, we plot the dominant mode of GW strain $h_{22}^+$ multiplied by the distance of the observer to origin $R$ from BNS mergers. The plot shows that in the first several orbits, the change in $n_{1/2}$ does not induce any significant difference because of the tidal effect, and equivalently, the EoS of each neutron star does not considerably play a role in the inspiral process. After about $10$ ms, the difference in the waveforms becomes conspicuous. The merger of the star from PCM with $n_{1/2} = 3n_0$ is delayed by approximately $5$ ms compared to that with $n_{1/2} = 2n_0$. In addition, one can see that the location of topology change affects the number of the inspiral orbits, i.e., the number of the peaks in the inspiral phase---the number of the peaks before merger is defined as the maximum of the amplitude of the GWs. Thus, it is easy to conclude that the larger $n_{1/2}$, the more is the number of the peaks. This is more distinct for the slightly lighter neutron star (Fig.~\ref{fig:GWa}). This is within the detection ability of the on-going and up-coming facilities, especially the ground-based facilities~\cite{GWDetection}. We expect that in the future, a more precise measurement of the GWs can give a hint regarding the location of the topology change, if it really exists.

From the simulations of the BNS mergers with heavier NSs, the plots shown in Fig.~\ref{fig:GWb} indicate that the merger occurs faster and eventuated in delayed black hole formation---the GWs disappear. In addition, the more similar is the process to the merger with lighter neutron stars, the greater is the delay in the merger process owing to the increase in $n_{1/2}$.

The position of the topology change also affects the speed of the evolution to black hole (see Fig.~\ref{fig:GWb}). When $n_{1/2}$ is larger, the evolution time is longer. This is because, the bigger $n_{1/2}$, the harder is the EoS in the core of the NS, and hence, it is more difficult to exchange the matter between the BNSs. This conclusion can also be extracted from the light BNS merger, as shown in Fig.~\ref{fig:GWa}. This figure shows that it takes more time to form a baby NS with an axisymmetric state for $n_{1/2} = 3 n_0$.

\begin{figure*}[tbp]
	\centering
	\includegraphics[width=1.0\linewidth]{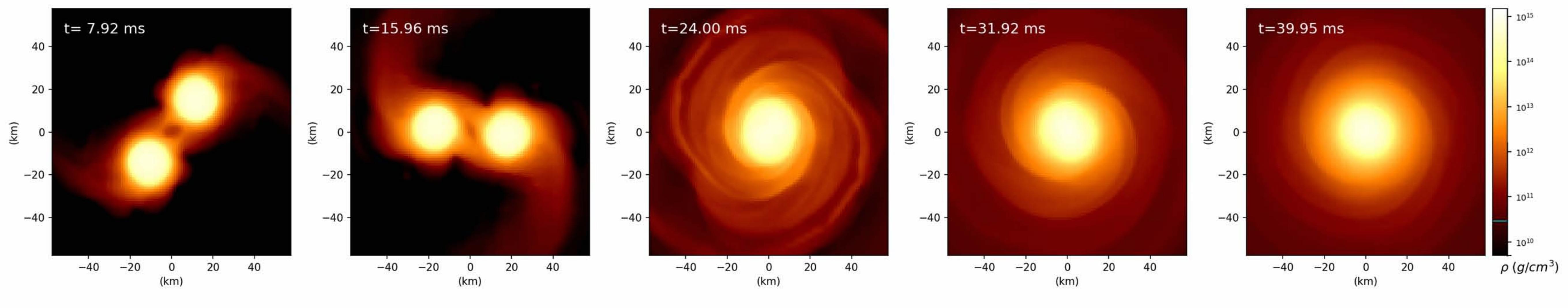}
    \includegraphics[width=1.0\linewidth]{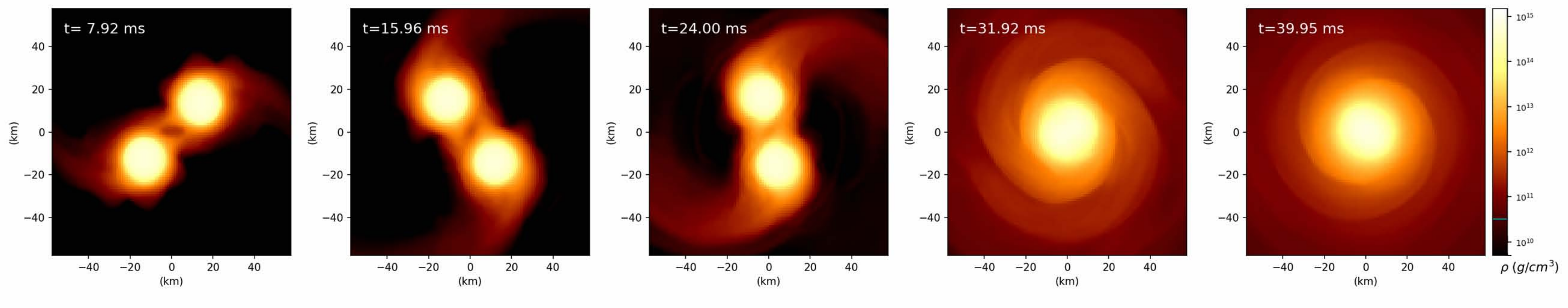}
	\caption{Matter density evolution of BNS mergers with equal mass $1.5M_\odot$ with (a) $n_{1/2}=2n_0$ (the upper row) and (b) $n_{1/2}=3n_{0}$ (the lower row).
}
	\label{fig:matter}
\end{figure*}

Finally, in Fig.~\ref{fig:matter}, we show the distribution of the matter evolution of the BNS merger with equal-mass neutron stars with $1.5M_\odot$ but different $n_{1/2}$. We found that the matter evolves faster when $n_{1/2} = 2n_0$ (the EoS is softer) than when  $n_{1/2} = 3n_0$ (the EoS is stiffer). Therefore, compared to the merger process with $n_{1/2} = 3n_0$, that with $n_{1/2}=2n_0$ is faster, and the stars merge easily and the inspiral period is shorter. This observation agrees with our expectation based on the waveform, as discussed above.

\section{Discussion and perspectives}

It is known that the topological structure that ubiquitously exists in highly correlated condensed matter can also induce novel phenomena in dense nuclear matter. The chiral effective model implemented the topology change at density $\geq 2n_{0}$, and the possible emergent hidden local flavor symmetry and scale symmetry in high-density nuclear matter yield the PCM for dense nuclear matter---the sound speed approaches the conformal limit, but the trace of the energy--momentum tenor is not zero---thus, all the constraints from terrestrial experiments and astrophysical observations are satisfied.

In addition to the theoretical prediction of the precocious appearance of the conformal sound speed, we simulated the experimentally interesting effect of topology change on the waveform of the GWs emitted from the BNS mergers discussed in this paper. We also demonstrated that the location of the topology change affects the number of inspiral orbits and the evolution time in the postmerger period more significantly than other global properties. Further, we demonstrated that the differences arising from the topology change are measurable in the on-going and up-coming facilities of GW detection. The physical mechanism behind these effects is that the stiffness of the EoS is changed by the change of $n_{1/2}$, and therefore, the efficiency of the matter exchange between the BNSs is affected. We can state explicitly that the larger $n_{1/2}$, the stiffer is the EoS, and therefore, the more difficult is the matter exchange.

The topology effect has been extensively studied and confirmed in condensed matter physics, but the topic has not received sufficient attention in the field of nuclear physics. The present work, in combination with other astrophysical observations, gives the first hint to observe or distinguish the topology effect in nuclear physics.

\acknowledgments

We thank all contributors of Einstein
Toolkit and LORENE code and thank Roberto De
Pietri and Parma University gravity group for modifying Einstein Toolkit and developing PyCactus code and making them publicly available. We would like to thank W. T. Ni for the valuable disusions we had. The work of YLM was supported in part by the National Science Foundation of China (NSFC) under Grant No. 11875147, No. 11475071 and the Intensive Study of Future Space Science Missions of the Strategic Priority Program on Space Science.
YLW is supported in part by NSFC under Grants No. 11851302, No. 11851303,
No. 11690022, No. 11747601, and the Strategic Priority Research Program of the Chinese
Academy of Sciences under Grant No. XDB23030100 as well as the CAS Center for Excellence in Particle Physics (CCEPP).

\end{document}